\documentclass[twocolumn,showpacs,preprintnumbers,amsmath,amssymb]{revtex4}

\usepackage{graphicx}
\usepackage{dcolumn}
\usepackage{bm}

\begin{document}


\title{Spontaneous spin textures in dipolar spinor condensates}
\author{S. Yi$^{1,2}$ and H. Pu$^1$}

\affiliation{$^1$Department of Physics and Astronomy, and Rice
Quantum Institute, Rice University, Houston, TX 77251-1892, USA}

\affiliation{$^2$Institute of Theoretical Physics, The Chinese
Academy of Science, Beijing, 100080, China}

\begin{abstract}
We have mapped out a detailed phase diagram that shows the ground
state structure of a spin-1 condensate with magnetic dipole-dipole
interactions. We show that the interplay between the dipolar and
the spin-exchange interactions induces a rich variety of quantum
phases that exhibit spontaneous magnetic ordering in the form of
intricate spin textures.
\end{abstract}

\date{\today}

\maketitle

Recent achievement of chromium condensate provides a platform for
exploring the effects of magnetic dipolar interaction in dilute
atomic Bose-Einstein condensates (BECs)~\cite{Cr}. On the
theoretical side, early studies on dipolar condensates
\cite{scalar, dipreview} concentrated on the scalar atomic BECs
where the dipole moments are assumed to be polarized by, e.g.,
external magnetic fields. To unlock the magnetic dipole moments,
one has to resort to optically confined spinor
condensates~\cite{dan,spinor}. A widely used and very powerful
theoretical tool in treating spinor condensates is the single mode
approximation (SMA) in which different spin states are assumed to
share the same spatial wave function \cite{yi1,law}. An important
implication, as well as a limitation, of the SMA is the lack of
spatially varying spin textures, i.e., the spins are uniformly
oriented under the SMA. The SMA is, however, expected to be valid
only when the total spin-dependent interaction (spin-exchange and
dipolar) is sufficiently weak compared to the spin-independent
interaction \cite{yi2}, and hence cannot cover the whole spectrum
of interesting quantum spin phenomena. The SMA becomes
particularly questionable in the presence of dipolar interactions
--- from the studies of ferromagnetic \cite{magdom} and
ferroelectric \cite{ivan} materials, it is known that even a
relatively weak dipolar interaction may spontaneously induce
spatially varying dipole moments, a fact which can be attributed
to the long-range anisotropic nature of the dipolar interaction.

In the present work, we investigate the ground state wave function
and spin structure of a dipolar spinor condensate by directly
minimizing the mean-field energy functional without the assumption
of the SMA. The main result is summarized in Fig.~\ref{phase}
where a phase diagram in the parameter space of the relative
dipolar strength and the trap aspect ratio is present. It can be
seen that the competition between the long-range dipolar and the
short-range exchange interactions gives rise to a rich variety of
spin textures. These spin textures emerge {\em spontaneously} in
the ground state, in contrast to the case of non-dipolar spinor
condensates, where spin textures can only be induced dynamically
\cite{yip,matt,stoof2}.

We consider an $F=1$ condensate with three magnetic sublevels
$m_F=1,0,-1$. In the mean-field treatment, the grand-canonical
energy functional of the system can be expressed as
\begin{eqnarray*}
E[\phi_\alpha] &=&\int d{\mathbf r}\left[\sum_\alpha\frac{\hbar^2
|\nabla\phi_\alpha|^2}{2m}-\mu n
+Vn+\frac{c_0n^2}{2}+\frac{c_2{\mathbf S}^2}{2}\right]\nonumber\\
&&+\frac{c_d}{2}\int \frac{d{\mathbf r}d{\mathbf r}'}{|{\mathbf
R}|^3}\left[{\mathbf S}({\mathbf r})\cdot{\mathbf S}({\mathbf
r}')-\frac{3{\mathbf S}({\mathbf r})\cdot{\mathbf R}\,{\mathbf
S}({\mathbf r}')\cdot{\mathbf R}}{|{\mathbf R}|^2}\right],
\end{eqnarray*}
where $\phi_\alpha({\mathbf r})=\sqrt{n_\alpha({\mathbf
r})}\,e^{i\Theta_\alpha({\mathbf r})}$ is the macroscopic wave
function of the $m_F=\alpha$ atoms with $n_\alpha$ being the
density and $\Theta_\alpha$ being the phase of the component
$\alpha$; $\mu$ is the chemical potential, $n=\sum_\alpha
n_\alpha$  the total density,
$V=m\omega_0^2(x^2+y^2+\lambda^2z^2)/2$ is trapping potential with
$\lambda$ being the trap aspect ratio, ${\mathbf
S}=\sum_{\alpha\beta}\phi^*_\alpha{\mathbf
F}_{\alpha\beta}\phi_\beta$ is the density of spin with ${\mathbf
F}$ being the angular momentum operator, and ${\mathbf R}={\mathbf
r}-{\mathbf r}'$. The collisional interactions include a
spin-independent part $c_0=4\pi\hbar^2(a_0+2a_2)/(3m)$ and a
spin-exchange part $c_2=4\pi\hbar^2(a_2-a_0)/(3m)$ with $a_f$
($f=0,2$) being the $s$-wave scattering length for two spin-1
atoms in the combined symmetric channel of total spin
$f$~\cite{spinor}. For the two experimentally realized spin-1
condensates $^{87}$Rb \cite{chapman} and $^{23}$Na \cite{kurn},
the spin-exchange interactions are ferromagnetic ($c_2<0$) and
anti-ferromagnetic ($c_2>0$), respectively. The strength of the
dipolar interaction is characterized by
$c_d=\mu_0\mu_B^2g_F^2/(4\pi)$ with $\mu_0$ being the vacuum
magnetic permeability, $\mu_B$ the Bohr magneton, and $g_F$ the
Land\'{e} g-factor.

The condensate wave function is found by numerically minimizing
the energy functional $E[\phi_\alpha,\phi_\alpha^*]$ subject to
the constraint of fixed total number of atoms $N$ using the
steepest descent method. In our calculation, we fix the value of
$c_0$ to be positive and choose $c_2/c_0=-0.01$ and $0.03$ for the
case of the ferromagnetic and anti-ferromagnetic spin-exchange
interaction, respectively. These two ratios correspond to the
scattering parameters of $^{87}$Rb and $^{23}$Na, respectively
\cite{kemp}. To focus on the dipolar effects, we allow $c_d$ to
vary and introduce $q\equiv c_d/|c_2|\geq0$ to measure the
relative strength of the dipolar interaction. The trap aspect
ratio $\lambda$ (which is varied between $0.1$ and $10$ in our
calculations) is another control knob that allows us to tune the
shape of the condensate and hence the effective dipolar
interaction \cite{dipreview, yi1}. Finally, in all numerical
results presented in this paper, we use $N=10^5$ and
$\omega_0=2\pi\times100$Hz.

Figure \ref{phase} summarizes the main result of this paper in the
form the phase diagram of the dipolar spin-1 condensate in the
$q$-$\lambda$ parameter space. For the SMA-I, II, and III phases,
the SMA is found to be valid and the spins are uniformly oriented
towards a direction determined by the dipolar interaction. In the
phases labelled by C, S, and P, the spatial wave functions are
spin-dependent and the system exhibits rich spin textures.

\begin{figure}
\centering
\includegraphics[width=3.3in]{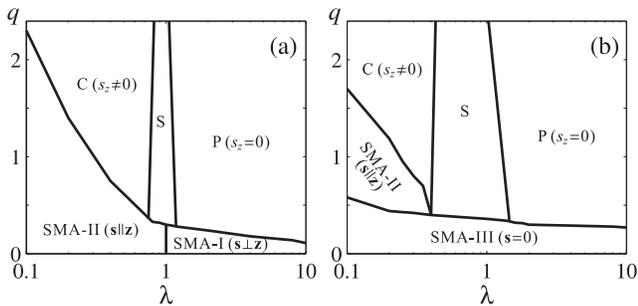}
\caption{Phase diagram of dipolar spin-1 condensates of
ferromagnetic (a) and anti-ferromagnetic (b) spin-exchange
interaction. ${\bf s}={\bf S}/n$ is the normalized local spin
vector.} \label{phase}
\end{figure}

{\em SMA ground states} --- The SMA implies that the normalized
local spin vector ${\mathbf s}\equiv{\mathbf S}/n$ is a constant.
For a ferromagnetic system in the absence of the dipolar
interaction, the vector ${\bf s}$ has a length of unity and the
ground state is degenerate about the orientation of $\mathbf s$
due to the SO(3) symmetry possessed by the order parameter
\cite{spinor}. When the dipolar interaction is present, the SO(3)
symmetry is broken and, as shown in Fig. \ref{phase}(a), the SMA
is valid only when $q$ is below a $\lambda$-dependent critical
value. In the SMA regions, the direction of ${\bf s}$ is
determined by the trapping geometry: ${\mathbf s}$ is
perpendicular to the $z$-axis for a pancake-shaped ($\lambda>1$)
condensate (SMA-I) and parallel to the $z$-axis for a cigar-shaped
($\lambda<1$) condensate (SMA-II), reminiscent of a quantum
ferromagnet with easy-plane and easy-axis anisotropy,
respectively. This result agrees with the quantum mechanical
calculation under the SMA~\cite{yi1}.

As shown in Fig. \ref{phase}(b), for an anti-ferromagnetic system,
the SMA is again valid for small $q$. When $q$ is below a critical
value that is rather insensitive to $\lambda$, the system exhibits
a vanishing spin with ${\mathbf s}=0$ (SMA-III). For $q$ above
this critical value and a cigar-shaped trapping potential with
$\lambda \lesssim 0.4$, the dipolar interaction may overwhelm the
spin-exchange interaction and make the condensate effectively
ferromagnetic, and the system enters the SMA-II phase with spin
oriented along the $z$-axis. We do not find in this case the SMA-I
phase with ${\bf s} \perp {\bf z}$. It is also worth mentioning
that transitions between different SMA phases are all first order.

Figure \ref{phase} shows that the critical dipolar strength at
which the SMA becomes invalid decreases as $\lambda$. For a
ferromagnetic system, $q_{\rm cr}=0.1$ at $\lambda=10$, which can
be reached in $^{87}$Rb~\cite{yi1}. The dipolar effects thus
become more prominent as the condensate becomes more
two-dimensional (2D) pancake-like. This is consistent with the
study in solid magnetic materials, where the dipolar interaction,
normally weak enough to be ignored in bulk materials, plays an
essential role in stabilizing long-range magnetic order in 2D
systems, e.g., in magnetic thin films \cite{film}. Our work shows
that it is feasible to observe non-SMA ground state in a pancake
$^{87}$Rb condensate.

\begin{figure}
\centering
\includegraphics[width=2.4in]{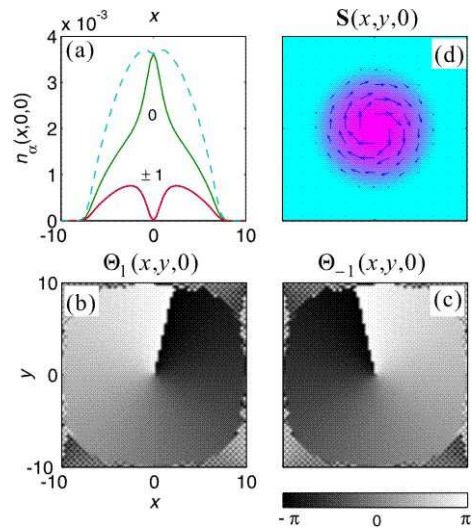}
\caption{(Color online). Ground state of the P phase for
ferromagnetic coupling with $\lambda=2$ and $q=1$. (a) Density of
each spin component along the $x$-axis. The total density is
denoted by the dashed line. (b) and (c) are the respective phase
images for $\phi_1$ and $\phi_{-1}$ in the $z=0$ plane. The phase
of $\phi_0$ is a constant and is not shown here. The phases in the
$z\neq 0$ plane are similar. (d) The vector plot of ${\mathbf S}$
in the $xy$-plane with the color map corresponding to the total
density. Here, as well as in Figs.~\ref{nsma2} and \ref{nsma3}, we
adopt $a_{\rm ho}=(\hbar/m\omega_0)^{1/2}$ and $N/a_{\rm ho}^3$ as
the units for length and density, respectively.} \label{nsma1}
\end{figure}

{\em Non-SMA ground states} --- The SMA regime is analogous to the
single-domain regime in micromagnetics, where the exchange energy
dominates over all other spin-dependent energy terms and a uniform
magnetization forms as a result \cite{magdom}. As $q$ increases,
the SMA eventually breaks down. At large $q$, the dipolar
interaction will dominate the spin-dependent interaction and the
difference between the ferromagnetic and anti-ferromagnetic
spin-exchange interaction becomes insignificant, as can be seen
from Fig.~\ref{phase}. The non-SMA region is further divided into
three distinct phases P, C and S, which stands for pancake, cigar
and spherical, respectively (see below).

Figure \ref{nsma1} shows a typical wave function in the P phase
which occurs in pancake geometry. The density profiles of all spin
components are axially symmetric with $n_1({\bf r})= n_{-1}({\bf
r})$ and hence $s_z=0$, i.e., the spins are planar and lie in the
$xy$-plane. The phases $\Theta_\alpha$ take the form
\begin{eqnarray}
\Theta_\alpha=w_\alpha\varphi+\varphi_\alpha,\label{phase}
\end{eqnarray}
where $\varphi$ is the azimuthal angle, $\varphi_\alpha$ is a
constant phase that satisfies
\begin{equation}
\varphi_1+\varphi_{-1}-2\varphi_0 = 0 \,, \label{relphase}
\end{equation}
and $w_\alpha$ is the phase winding number with the values
\[\langle w_1,w_0,w_{-1}\rangle=\langle-1,0,1\rangle\,.\]
Therefore the two spin components, $\phi_1$ and $\phi_{-1}$, are
in vortex states with opposite winding numbers. Due to the
presence of the non-rotating $\phi_0$ component, the total density
does not vanish at the vortex core. Such a ground state represents
a coreless skrymion. It is easy to show that the spin vector can
be expressed as
\begin{eqnarray}
{\mathbf S}=2\sqrt{2n_0
n_1}\,(\sin\varphi,\;-\cos\varphi,\;0)\,.\nonumber
\end{eqnarray}
As illustrated in Fig. \ref{nsma1}(d), the spin continuously curls
around the $z$-axis to form a vortex such that the system
possesses a persistent spin current but no net density current.
The spin vortex state is analogous to the flux-closure magnetic
states in micromagnetics \cite{magdom}. In the vicinity of the
$z$-axis, to reduce the exchange energy, the magnitude of the spin
gradually decreases and vanishes on the $z$-axis. This differs
from the magnetic vortex observed in nanoscale ferromagnets
\cite{buss} where, due to the conservation of local spin moment,
the magnetization in the vortex core turns toward the $z$-axis
\cite{magdom}.

\begin{figure}
\centering
\includegraphics[width=2.5in]{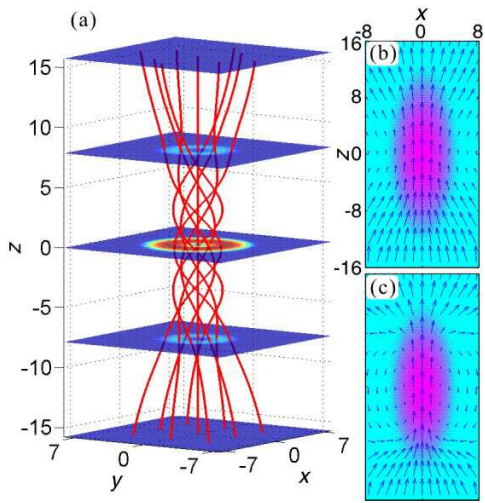}
\caption{(Color online). (a) The streamline plot of the spins for
ferromagnetic coupling in the C phase with $\lambda=0.4$ and
$q=1.6$. The color map represents $|{\mathbf S}_\perp|$ on these
cross-sections. (b) The vector plot of ${\mathbf s}_\parallel$ in
the $xz$-plane. Same parameters as in (a). (c) The same as (b)
except for $q=6$.} \label{nsma2}
\end{figure}

In the C phase which occurs in a cigar-shaped trap, $n_\alpha$
remains to be cylindrically symmetric and $\Theta_\alpha$ can also
be expressed in the form of Eq.~(\ref{phase}) where the phases
$\varphi_\alpha$ are now $z$-dependent but still satisfy
Eq.~(\ref{relphase}), and the winding numbers are now given by
\[\langle w_1,w_0,w_{-1}\rangle=\langle 0,1,2\rangle\,.\] The spin
vector in the C phase takes the form
\begin{eqnarray}
{\mathbf S}=\big[\Delta \sin(\varphi+\delta),\; -\Delta
\cos(\varphi+\delta),\;n_1-n_{-1}\big]\,,\nonumber
\end{eqnarray}
where $\Delta \equiv \sqrt{2n_0}(\sqrt{n_1}+\sqrt{n_{-1}})$, and
$\delta(z)\equiv\varphi_0(z)-\varphi_1(z)$ is the spin twisting
angle which is a monotonically increasing function of $z$ with
$\delta(z=0)=0$. A typical spin texture plot in the C phase is
shown in Fig.~\ref{nsma2}(a) where one can see that the spins
twist around the $z$-axis, tracing out a helical pattern. The
total twisting angle $\delta(\infty)-\delta(-\infty)$ increases
with $q$ and approaches $\pi$ at large $q$. The corresponding
structures of the planar spin ${\mathbf
S}_\parallel\equiv(S_x,S_z)$ are plotted in Fig. \ref{nsma2}(b)
and (c), which resembles the magnetization of a bar magnet. Note
that, in the absence of the spin-exchange interaction ($c_2=0$),
the C phase does not exist and is replaced by the SMA-II phase
with spins uniformly oriented along $z$.

%

\begin{figure}
\centering
\includegraphics[width=2.5in]{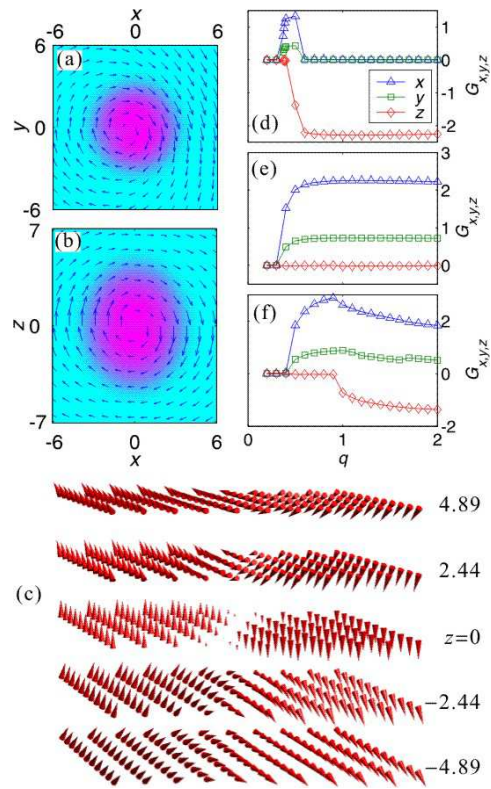}
\caption{(Color online). The spin structure in the S phase for
anti-ferromagnetic coupling. (a) Spin projection in the $z=2.29$
plane. $\lambda=0.4$ and $q=1$. (b) Spin projection in the $y=0$
plane. $\lambda=0.8$ and $q=1$. (c) The spin structure for
$\lambda=0.8$ and $q=1$. (d)-(f) The dipolar strength dependence
of the toroidal moments for $\lambda=1.2$, $0.8$, and $0.4$,
respectively.} \label{nsma3}
\end{figure}

In between the P and C phases, is the S phase which occurs when
the trapping potential is close to spherical ($\lambda \approx
1$). A distinct feature of the S phase is that $n_\alpha$ becomes
non-axisymmetric, signalling the broken of the cylindrical
symmetry of the spatial wave functions. To further quantify the
spin texture, we define the spin toroidal moment vector as
\cite{ivan}
\[{\mathbf G}\equiv\int d{\mathbf r}\,\left[{\mathbf
r}\times{\mathbf S}({\bf r})\right]\,.\]For all the SMA phases, we
have ${\bf G}=0$. For both the P and C phases, ${\bf G}$ points
along the $z$-axis. The S phase, by contrast, features finite
$G_x$ and $G_y$. The toroidal moments as functions of $q$ is
plotted in Fig.~\ref{nsma3}(d)-(f) for three different values of
$\lambda$. For small $q$, ${\bf G}=0$ and the system is in the SMA
regime. As $q$ exceeds a threshold, ${\bf G} \neq 0$ and the S
phase is reached. The continuity of ${\bf G}$ suggests, in this
mean-field calculation, the transitions between SMA and non-SMA,
and also among non-SMA, phases are of second order. In
Fig.~\ref{nsma3}(d), we also see that for $q>0.6$, the system
enters the P phase. Fig.~\ref{nsma3}(a)-(c) show examples of the
spin structure in the S phase. For this particular set of
parameters, the condensate displays two domains: one has $S_z>0$
and and the other $S_z<0$.

{\em Collapse} --- Condensate collapse will occur when the dipolar
interaction strength exceeds a critical value $c_d^*$. For both
ferromagnetic and anti-ferromagnetic coupling, we have found that
$c_d^*\approx0.24c_0$ and is nearly independent of the trap aspect
ration $\lambda$. This critical value agrees with that of a highly
elongated dipolar scalar condensate with dipole moments polarized
along the axial direction \cite{scalar}. As is well known, the
critical dipolar strength for scalar condensate is very sensitive
to $\lambda$ \cite{scalar}. In the spinor system studied here, the
dipoles are free to rearrange themselves to minimize the dipolar
interaction, which renders the insensitivity of $c_d^*$ with
respect to the trapping geometry.

{\em Dipole induced spin-orbital coupling}
--- In the SMA regime, the total orbital angular
momentum {\bf L} vanishes, and the ground state wave functions
$\phi_\alpha$ can be taken to be real. In the non-SMA regime, the
spin and orbital degrees of freedom are intimately coupled and
$\phi_\alpha$ must be described by complex functions. Here the
spin (${\bf S}^2$) and orbital (${\bf L}^2$) angular momenta are
not separately conserved, only the total angular momentum $({\bf
S}+{\bf L})^2$ is conserved. It is not difficult to show that, in
the non-SMA regime, if the set of wave functions $\{\phi_\alpha\}$
minimizes the energy functional, then the set $\{\phi_\alpha^*\}$
does not. On the other hand, if we make the transformation: $
(\phi_1, \phi_0, \phi_{-1}) \rightarrow (\phi_{-1}^*, -\phi_0^*,
\phi_{1}^*)$, then the new set still minimizes the energy. The
above transformation amounts to inverting ${\bf S}$ and ${\bf L}$
{\em simultaneously} and hence preserving the relative orientation
between them. This inter-connection between ${\bf S}$ and ${\bf
L}$ is well known in superfluid $^3$He and the related phenomenon
has been termed the spontaneously broken spin-orbit symmetry
\cite{he3,legg}.

The dipole induced spin-orbital coupling is recently explored
theoretically in the dynamics of $^{52}$Cr condensate and is
predicted to manifest in the Einstein-de Haas effect \cite{haas}.
$^{52}$Cr atom features a spin-3 ground state \cite{diener} whose
short-range collisional interaction is characterized by four
scattering lengths and not all of them are accurately known.
Although we have studied, in this work, a relatively simpler
spin-1 system, we expect much of the essential physics can be
applied to higher spin systems.

In conclusion, we have provided a detailed phase diagram which
shows the ground state structure of a dipolar spin-1 condensate
beyond the SMA. In this system, the spin and orbital degrees of
freedom are intimately coupled together. The interplay between the
long-range dipolar and the short-range exchange interactions gives
rise to a variety of quantum phases characterized by distinctive
spin textures. As such, dipolar spinor condensates represent an
intriguing quantum magnetic system whose properties are highly
tunable. The work here assumes zero magnetic field, an assumption
valid for $^{87}$Rb and $^{52}$Cr when the external magnetic field
strength is less than $0.1\,$mG \cite{yi1}. This constraint poses
an experimental challenge but is definitely within the reach with
current technology.

We thank Profs. Michael Chapman, Jason Ho, Carl Rau and Li You,
and Dr. Jian Li for many insightful discussions. HP acknowledges
the hospitality of the Aspen Center for Physics, and support from
ORAU and NSF.

\end{document}